\documentstyle[12pt,aaspp4]{article}
\begin{document}

\title{Using the DIRBE/IRAS All-Sky Reddening Map To Select
Low-Reddening Windows Near the Galactic Plane}

\author{K. Z. Stanek\altaffilmark{1}} 
\affil{Harvard-Smithsonian Center for Astrophysics, 
60 Garden St., MS20, Cambridge, MA 02138}
\affil{e-mail: kstanek@cfa.harvard.edu} 
\altaffiltext{1}{On leave from Copernicus Astronomical Center, 
Warsaw, Poland}

\begin{abstract}

Recently Schlegel, Finkbeiner \& Davis published an all-sky reddening
map based on the COBE/DIRBE and IRAS/ISSA infrared sky surveys.  Using
the reddening map of Baade's Window and sample of 19 low-latitude
($|b|<5\deg$) Galactic globular clusters I find that the DIRBE/IRAS
reddening map overestimates $E(B-V)$ at low galactic latitudes by a
factor of $\sim 1.35$. I also demonstrate the usefulness of this high
resolution map for selecting low-reddening windows near the Galactic
plane.

\end{abstract}


\section{INTRODUCTION}

We live in a dusty Universe (Hoover 1998, private communication), and
correcting for the dust extinction and reddening affects almost all
aspects of optical astronomy. For us, observing from within the Milky
Way, it is of crucial importance to know how much Galactic dust there
is towards various objects. Burstein \& Hailes (1982; hereafter: BH)
constructed an all-sky reddening map, used extensively by the
astronomical community.\footnote{Their paper was cited 540 times
between 1992 and 1997} Recently, Schlegel, Finkbeiner \& Davis (1998;
hereafter: SFD) published a new all-sky reddening map, based on the
COBE/DIRBE and IRAS/ISSA maps.\footnote{The reddening map and related
files and programs are available using the {\tt WWW} at: {\tt
http://astro.berkeley.edu/davis/dust/}} This map is intended to
supersede the BH map in both the accuracy and the spatial resolution
($6.1'$). Also, unlike the BH map, which presented the $E(B-V)$ values
only for the $|b|>10\deg$ regions, the SFD map extends all the way to
the Galactic plane.

There are many instances when one would like to be able to identify
low-reddening regions near the Galactic plane. As an example, the
expected microlensing optical depth increases strongly towards the
Galactic center (e.g. Stanek et al.~1997, their Figure~12).  In this
paper I show the usefulness of the SFD map for selecting low-reddening
regions close to the Galactic plane.  In Section~2 I compare the
predicted and the observed reddenings and arrive at the approximate
scaling relation. In Section~3 I show the re-scaled reddening map for
central region of our Galaxy and discuss its applications for
microlensing experiments.

\section{COMPARING THE REDDENING VALUES}

With its high spatial accuracy, the SFD map is potentially very useful
for selecting low-reddening windows close to the Galactic plane.
However, as discussed by SFD, at low Galactic latitudes ($|b|<5\deg$)
most of the contaminating sources have not been removed from their
map.  Also, no comparisons between the predicted reddenings and
observed reddenings have been made in these regions. It is therefore
important to attempt such a comparison.

Harris (1996) surveyed the vast literature on the Galactic globular
clusters (GCs), producing an electronic catalog\footnote{The catalog
is available using the {\tt WWW} at: {\tt
http://www.physics.mcmaster.ca/Globular.html}} of GCs with reasonably
well-known properties, among them the reddening $E(B-V)$.  Many
Galactic globular clusters are extensively studied objects, for which
it is possible to accurately determine their extinction and reddening
using a variety of methods (Burstein \& Heiles 1978; Webbink 1985;
Zinn 1985; Reed, Hesser \& Shawl 1988). However, many GCs in Harris
(1996) compilation have been poorly studied, in which cases the
reddening estimates are rather inaccurate.

There are 147 GCs in the catalog of Harris (1996), out of which I
selected 32 clusters with $|b|<5\deg$. I then excluded eight clusters
with the distance from the Galactic plane $|z|<0.3\;kpc$, to ensure
that each cluster remaining in the sample probes most of the
reddening along the line of sight (see discussion in Stanek~1998).
For the remaining 24 clusters I used the program ``dust\_getval''
included with the SFD reddening map to obtain the value of reddening
predicted by the SFD, which I hereafter call $E(B-V)_{SFD}$. This was
done by obtaining 49 values of $E(B-V)_{SFD}$ on a uniform $7\times7$
grid with $2.5'$ spacing centered on the cluster ($15'\times15'$ part
of the sky), which allowed me to obtain an average value of
$E(B-V)_{SFD}$ as well as its standard deviation $\sigma_{E(B-V)}$.
At this point I excluded additional five GCs with
$\sigma_{E(B-V)}>0.2\;$mag, indicating large reddening gradient across
the cluster (Stanek 1998), which might strongly affect the determined
reddening.  This leaves me with 19 GCs with which I can test the SFD
map for the $|b|<5\deg$ region.  I denote the GCs reddenings taken
from the catalog of Harris by $E(B-V)_{GC}$.

\begin{figure}[t]
\plotfiddle{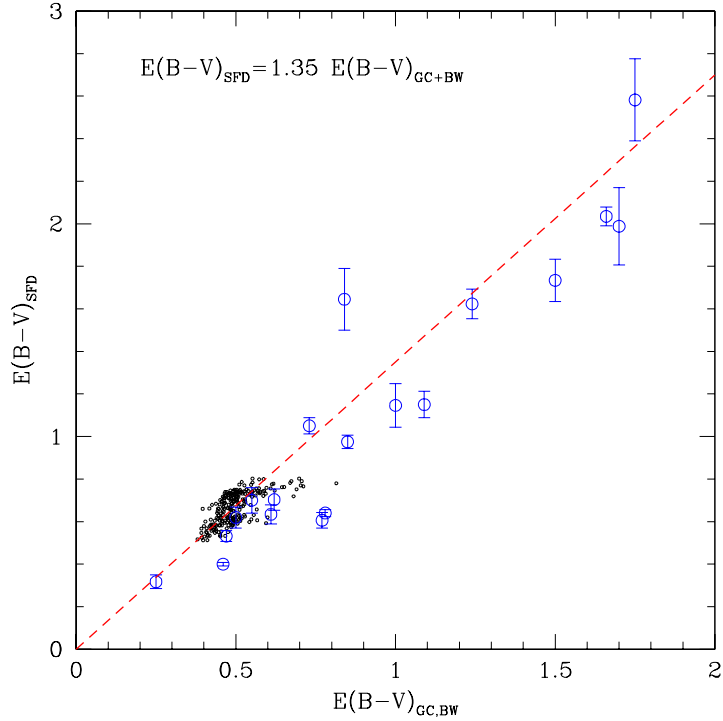}{8cm}{0}{50}{50}{-160}{-90}
\caption{Relation between the observed values of the reddening
$E(B-V)_{GC+BW}$, obtained using 19 globular clusters with $|b|<5\deg$
from catalog of Harris (1996) and the reddening map of Baade's Window
(Stanek 1996), and the predicted reddening $E(B-V)_{SFD}$ obtained
using the all-sky map of Schlegel, Finkbeiner \& Davis~(1998).  Large
symbols denote the clusters and small symbols denote points from
Baade's Window (for discussion see text). Dashed line indicates the
adopted relation $E(B-V)_{SFD}=1.35 E(B-V)_{GC+BW}$.}
\label{fig:sfd}
\end{figure}

Additional comparison comes from using the reddening map of Baade's
Window (BW) region (Stanek 1996). This map was constructed using the
method of Wo\'zniak \& Stanek (1996) and the OGLE project data
(Szyma\'nski \& Udalski 1993; Udalski et al.~1993). The zero point of
Baade's Window reddening map was re-determined recently by Gould,
Popowski \& Terndrup~(1998) and Alcock et al.~(1998). I obtained 240
values of $E(B-V)_{SFD}$ on a uniform grid with $2.5'$ spacing,
falling within the BW map.  For the same grid I also obtained the
values of $E(V-I)_{BW}$ using my map of the reddening. Using the
relations between the selective extinction coefficients given by SFD
(their Table~6) and also the coefficient of $A_V/E(V-I)=2.49$ found by
Stanek (1996), I then converted the $E(V-I)_{BW}$ reddening to
$E(B-V)_{BW}$ using relation $E(B-V)_{BW}=E(V-I)_{BW}/1.3$.

In Figure~\ref{fig:sfd} I plot the relation between the observed
values of the reddening $E(B-V)_{GC+BW}$, obtained using globular
clusters with $|b|<5\deg$ from catalog of Harris (1996) and the
reddening map of Baade's Window (Stanek 1996), and the predicted
reddening $E(B-V)_{SFD}$ obtained using the SFD map.  Large symbols
denote the clusters, with $\sigma_{E(B-V)}$ as errorbars, and small
symbols denote points from Baade's Window.  The observed reddenings
$E(B-V)_{GC}$ extend from 0.25 to 1.75. Using linear least-squares fit
to both the GCs and the BW samples I obtained the scaling relation
$E(B-V)_{SFD}=1.36 E(B-V)_{GC+BW}-0.02$, while using only the GCs I
obtained a relation $E(B-V)_{SFD}=1.50 E(B-V)_{GC}-0.26$.  Since the
typical error of the $E(B-V)_{BW}$ values is $\sim 0.05\;$mag (Stanek
1996), I feel that using both samples gives a better estimate of the
scaling relation, so hereafter I adopt $E(B-V)= E(B-V)_{SFD}/1.35$ as
representing the ``true'' reddening for the $|b|<5\deg$ region. It
should be stressed here that there is no physical reason for {\em one}
``universal'' relation of the kind fitted above, so the adopted
scaling factor of 1.35 should be applied carefully, only when there
are no additional constraints on the reddening.  Also, as the
comparison range of the $E(B-V)_{GC}$ extends only to 1.75, the
derived scaling might not apply in high reddening regions very close
to the Galactic plane ($|b|<2\deg$).  Finally, for regions with
$|b|>5\deg$ this scaling factor {\em should not} be applied (SFD;
Stanek 1998), as the SFD map is cleaned from the point sources in
these regions.

\section{REDDENING NEAR THE GALACTIC PLANE}

\begin{figure}[t]
\plotfiddle{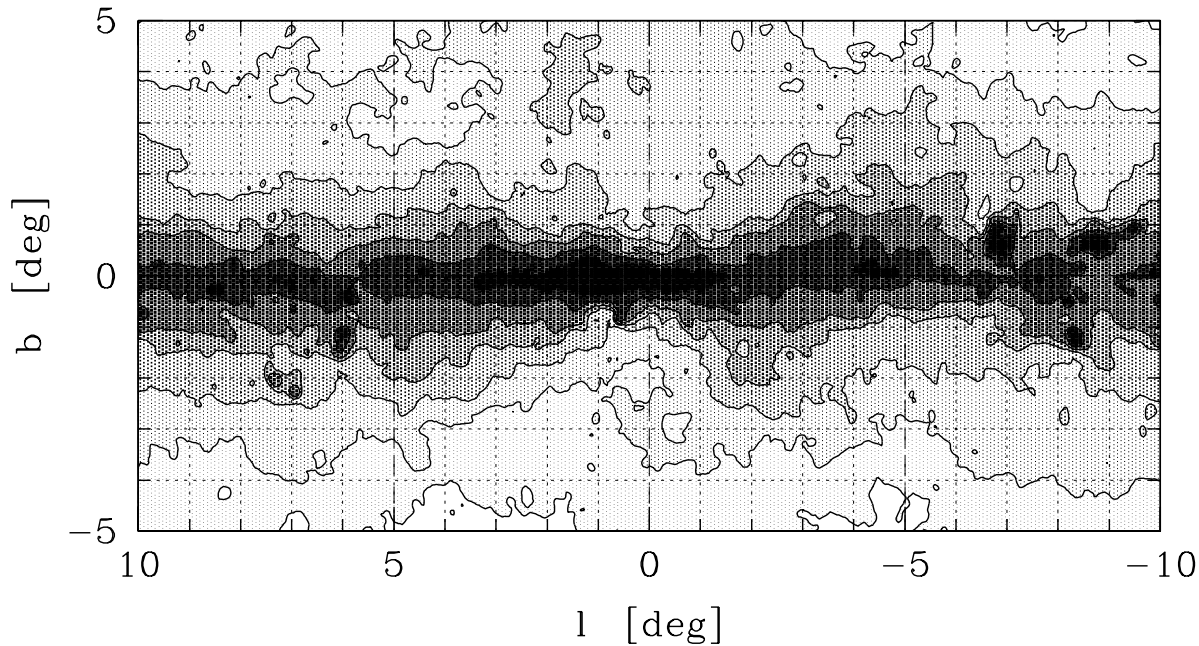}{7cm}{0}{80}{80}{-250}{-320}
\caption{Reddening map of $20\times10\;\deg$ part of the the Galactic
center. Contours correspond to re-scaled $E(B-V)=0.4, 0.8, 1.5, 3.0,
5.0$ and $10.0 \;$mag.}
\label{fig:dust}
\end{figure}

\begin{figure}[t]
\plotfiddle{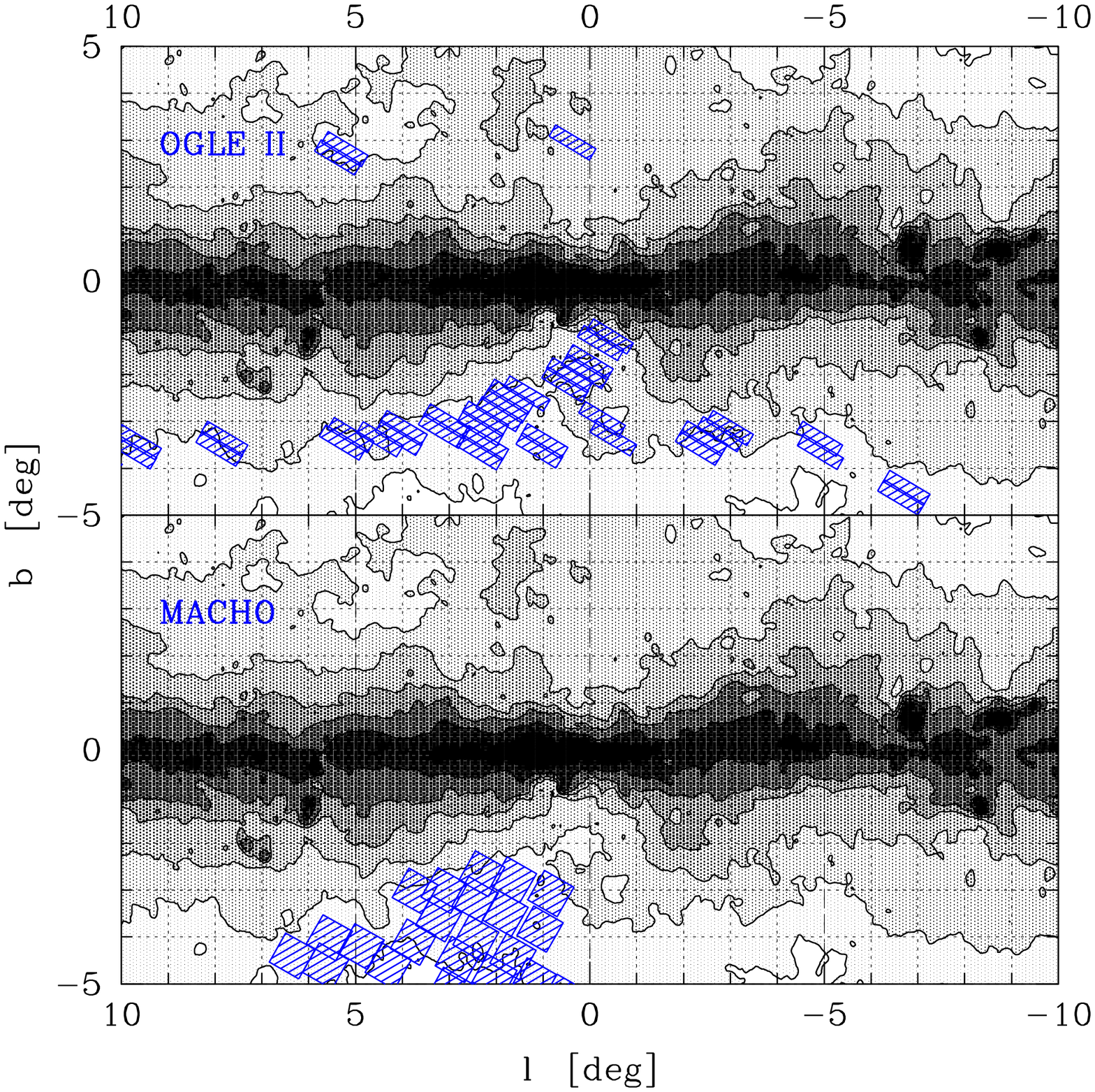}{10cm}{0}{60}{60}{-193}{-105}
\caption{Locations of the fields observed towards the Galactic bulge
by the OGLE II (upper panel) and by the MACHO (lower panel)
microlensing experiments.}
\label{fig:micro}
\end{figure}

I can now apply the SFD map and the scaling factor obtained in the
previous Section to investigate the reddening close to the Galactic
plane. In Figure~\ref{fig:dust} I plot the $E(B-V)$ reddening map of
$20\times10\;\deg$ part of the the Galactic center. Contours
correspond to re-scaled $E(B-V)=0.4, 0.8, 1.5, 3.0, 5.0, 10.0 \;$mag.
As discussed above, the comparison range of the $E(B-V)_{GC}$ extends
only to 1.75, so the derived scaling of $\sim 1.35$ might not apply in
high reddening regions very close to the Galactic plane ($|b|<2\deg$).
Also, the SFD map gives by its construction the total dust reddening
along given line of sight, which for regions close to the Galactic
plane might have significant contribution from dust distributed beyond
the observed source.

The spatial structure of reddening seen in Figure~\ref{fig:dust} is
very rich, both on large and small scales. Some of the features, such
as more dust extinction above the Galactic plane than below the plane,
are well known and can be observed even with a naked eye. There is a
pronounced ``pinch'' in the reddening at the $l\sim 0$ longitude,
where some of the iso-reddening contours are $1-3\deg$ closer to the
Galactic plane than elsewhere in the map. In a low-reddening window
near $(l,b)=(0,-2)$ the rescaled reddening is as low as $E(B-V)\sim
0.6\;$mag.  Also worth noticing is low-reddening region near
$(l,b)=(4,3)$ with the rescaled reddening as low as $E(B-V)\sim
0.55\;$mag.

It is interesting to compare the locations of the fields observed
towards the Galactic bulge by the OGLE II (Figure~\ref{fig:micro},
upper panel) and by the MACHO (Figure~\ref{fig:micro}, lower panel)
microlensing experiments (Udalski, Kubiak \& Szyma\'nski 1997; Alcock
et al.~1997). Both experiments cover similar areas on the sky ($\sim
12\deg^2$) and show significant overlap with each other, with OGLE II
taking somewhat better advantage of low-extinction regions close to
the Galactic center (such as the window near $(l,b)=(0,-2)$ discussed
above). However, OGLE II fields are much better distributed for
applications of their data for the Galactic structure studies, such as
investigating the properties of the Galactic bar with the red clump
giants (Stanek et al.~1997).

\begin{figure}[t]
\plotfiddle{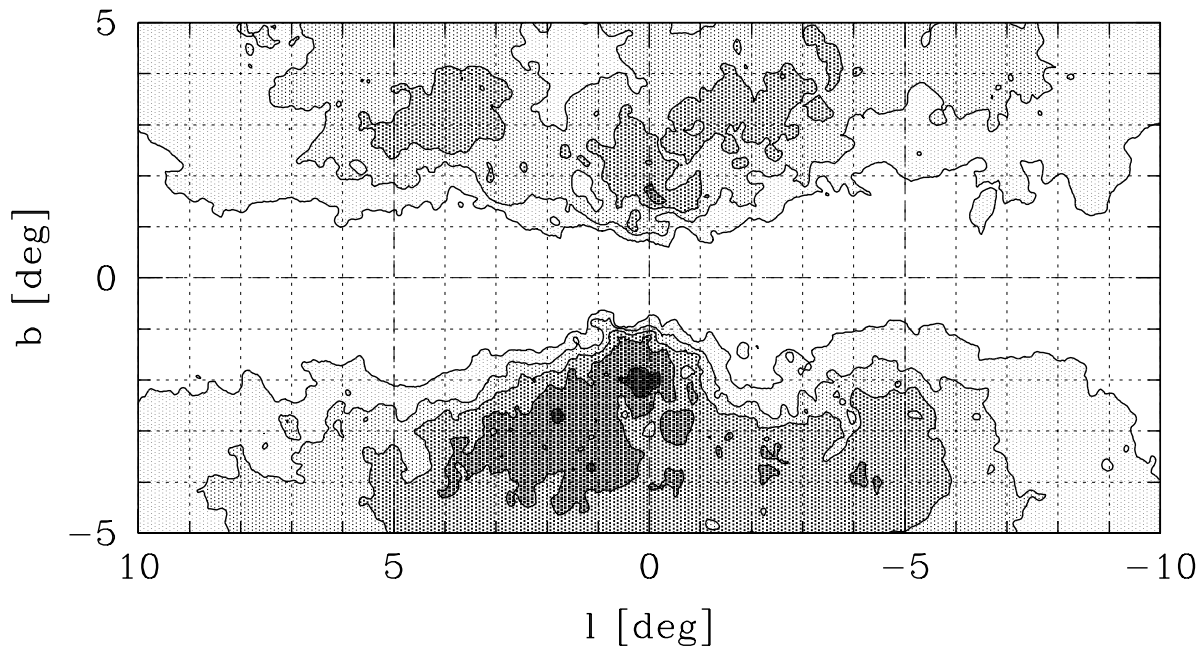}{7cm}{0}{80}{80}{-250}{-320}
\caption{Map of the ``reddening-adjusted optical depth''
$\tau_{E(B-V)}$ (Equation~1).  Contours are plotted at 10, 30, 50, 70
and 90\% of the peak value of $\tau_{E(B-V)}$.}
\label{fig:tau}
\end{figure}

We might attempt to define an ``optimal'' selection of fields to be
used for the microlensing studies. This selection naturally depends on
the scientific goal of a given study. To maximize the observed
microlensing optical depth during the lifetime of the experiment, I
define ``reddening-adjusted optical depth'' as
\begin{equation}
\tau_{E(B-V)}\equiv \tau \exp{[-\kappa E(B-V)]},
\end{equation}
where $\tau$ is the standard optical depth for microlensing and the
factor $\exp{[-\kappa E(B-V)]}$ is used to reduce $\tau$ in regions of
high reddening (and hence extinction). The coefficient $\kappa$ in the
exponent can be adjusted to reflect, for example, a photometric band
used for observations.  To give an example, the microlensing optical
depth is highest towards the Galactic center, yet because of the high
dust extinction there it is not a good place to monitor
microlensing. In Figure~\ref{fig:tau} I plot $\tau_{E(B-V)}$, using
the rescaled SFD map as the source of reddening $E(B-V)$ and the
optical depth from the bar model E2 of Stanek et al.~(1997, their
Figure~12) as the source of $\tau$ (the contribution from the Galactic
disk to the microlensing was not included). I took $\kappa=1.0$ for
simplicity.

In practice, as current microlensing experiments operate in
crowding-limited mode, the reddening-adjusted optical depth defined
above is not optimal from their point of view. However, with the
progressing development of image-subtraction techniques (see recent
papers by Alard \& Lupton 1998 and Tomaney 1998) one will be able to
detect the microlensing events regardless of crowding.

To summarize, the SFD map overestimates the reddening in $|b|<5\deg$
regions by a factor of $\sim 1.35$, but it is a very valuable tool in
investigating features in the reddening near the Galactic plane.  I
limited myself to investigating the central part of our Galaxy, but
the SFD map should be also useful for similar studies in other parts
of the Galactic plane.

\acknowledgments{I was supported by the Harvard-Smithsonian Center for
Astrophysics Fellowship.  Bohdan Paczy\'nski provided me with helpful
comments. As usual, the SM graphics package written by Robert Lupton
was most useful.}


\begin{references}

\reference{} Alard, C., \& Lupton, R. H., 1998, ApJ, submitted (astro-ph/9712287)

\reference{} Alcock, C., et al., 1997, ApJ, 479, 119

\reference{} Alcock, C., et al., 1998, ApJ, 494, 396

\reference{} Burstein, D., \& Hailes, C., 1982, AJ, 87, 1165 [BH]

\reference{} Gould, A., Popowski, P., \& Terndrup, D. M., 1998, ApJ, 492, 778

\reference{} Harris, W. E., 1996, AJ, 112, 1487

\reference{} Reed, B. C., Hesser, J. E., \&  Shawl,  S. J., 1988, PASP, 100, 545

\reference{} Schlegel, D. J.,  Finkbeiner, D. P., \&  Davis, M., 1998, ApJ,
	in press (astro-ph/9710327) [SFD]

\reference{} Stanek, K. Z., 1996, ApJ, 460, L37 

\reference{} Stanek, K. Z., et al., 1997, ApJ, 477, 163

\reference{} Stanek, K. Z., 1998, ApJ, submitted (astro-ph/9802093)

\reference{} Szyma\'nski, M., \& Udalski, A., 1993, AcA, 43, 91

\reference{} Tomaney, A. B., 1998, ApJ, submitted (astro-ph/9801233)

\reference{} Udalski, A., Szyma\'nski, M., Ka\l u\.zny, J., Kubiak, M.,
	\& Mateo, M., 1993, AcA, 43, 69

\reference{} Udalski,  A., Kubiak, M.,  \& Szyma\'nski, M., 1997,  AcA, 47, 319

\reference{} Webbink, R. F., 1985, in ``Dynamics of Star Clusters'',
        IAU Symposium 113, eds. J. Goodman and P. Hut (Dordrecht: Reidel), 541

\reference{} Wo\'zniak, P. R., \& Stanek, K. Z., 1996, ApJ, 464, 233 

\reference{} Zinn, R., 1985, ApJ, 293, 424

\end{references}
\end{document}